%% file: main.tex
\documentclass[%
article,
superscriptaddress,
 amsmath,amssymb,
 aps, 
]{revtex4-2}
\usepackage[colorlinks=true,linkcolor=blue]{hyperref}

\usepackage{subfigure}
\usepackage{xspace}
\usepackage{graphicx}
\usepackage{float}
\usepackage{amsmath}
\usepackage{tikz}
\usepackage{tikz-feynman}
\usetikzlibrary{shapes,automata,arrows,positioning,calc}
\usetikzlibrary{arrows.meta,bending}
\usepackage{amssymb}
\usepackage{comment}
\usepackage{xcolor}

\input{macros}

\newcommand{\Propagator}{\mathcal{G}}
\newcommand{\friction}{\mu}
\newcommand{\effeD}{D_{\text{eff}}}
\newcommand{\effeV}{\Omega}
\newcommand{\Hermite}[2]{H\! e_{#1}\left(#2\right)}
\newcommand{\rotDiffusion}{D_\text{r}}

\newcommand{\operator}{\mathcal{L}}
\newcommand{\Expp}[1]{\text{e}^{#1}}
\newcommand{\propga}[2]{G(#1,#2
)}
\newcommand{\rescalefric}{\beta}
\newcommand{\rescalefricTwoD}{\Tilde{\beta}}
 
\newcommand{\mass}{r}

\usetikzlibrary{decorations.pathmorphing}
\usetikzlibrary{decorations.markings}
\usetikzlibrary{snakes}
\tikzset{
xxtsubstrate/.style={decorate, 
line width=1pt,
draw=olive, 
decoration=snake, 
segment amplitude=0.75mm, 
line after snake=0.25mm,
line before snake=0.25mm
},
tsubstrate/.style={decorate, 
line width=1pt,
draw=olive, 
decoration=snake, 
segment amplitude=0.5mm, 
segment length=5pt,
segment amplitude=0.2mm, 
line after snake=1mm,
line before snake=1mm
},
Bsubstrate/.style={decorate, 
line width=1pt,
draw=olive, 
decoration=snake,
segment length=5pt,
segment aspect=0,
segment amplitude=0.5mm, 
line after snake=0mm,
line before snake=0mm
},
substrate/.style={decorate, 
line width=1pt,
draw=olive, 
decoration=snake, 
segment length=5pt,
segment amplitude=0.5mm, 
line after snake=0.5mm,
line before snake=0.5mm
},
activity/.style={very thick,draw=red,postaction={decorate},
decoration={markings,mark=at position .5 with
{\arrow[draw=red]{>}}}},
tactivity/.style={thick,draw=red,postaction={decorate},
decoration={markings,mark=at position .5 with
{\arrow[draw=red]{>}}}},
tEPSactivity/.style={thick,draw=red,postaction={decorate},
decoration={markings,mark=at position .55 with
{\arrow[draw=red]{>}}}},
tAactivity/.style={thick,draw=red},
Aactivity/.style={very thick,draw=red},
ghostactivity/.style={very thick,draw=white},
Bactivity/.style={very thick,draw=blue,dashed},
Baractivity/.style={very thick,draw=black},
Cactivity/.style={very thick,draw=cyan,snake},
tSactivity/.style={thick,draw=red,postaction={decorate},
decoration={markings,mark=at position .7 with
{\arrow[draw=red]{>}}}},
Sactivity/.style={very thick,draw=red,postaction={decorate},
decoration={markings,mark=at position .7 with
{\arrow[draw=red]{>}}}},
ABPactivity/.style={very thick,draw=red,decorate,decoration={coil,segment length=6pt}},
Arrowactivity/.style={decoration={markings,mark=at position 1 with
    {\arrow[scale=1.6,>=stealth]{>}}},postaction={decorate}},
triangle/.style = {fill=blue!10, regular polygon, regular polygon sides=3,minimum size=3mm},
    node rotated/.style = {rotate=180},
    border rotated/.style = {shape border rotate=180},}

\newcommand{\bareproplong}[2]{\tikz[baseline=-2.5pt]{
\draw[Aactivity] (180:1.2) -- (0,0)  node[at start ,above]{$#1$} ;
\draw[Aactivity] (0:0.6) -- (0,0) node[at start,above] {$#2$};}}
\newcommand{\bareprop}[2]{\tikz[baseline=-2.5pt]{
\draw[Aactivity] (180:0.6) -- (0,0)  node[at start ,above]{$#1$} ;
\draw[Aactivity] (0:0.6) -- (0,0) node[at start,above] {$#2$};}}

\newcommand{\barepropX}[4]{\tikz[baseline=-2.5pt]{
\draw[Aactivity]   (-0.8,0) --(0,0)   node[at start, above] {$#1$}  node[at start, below] {$#2$};
\draw[Aactivity] (0.8,0) -- (0,0) node[at start,above] {$#3$}
node[at start, below] {$#4$};
}}

\newcommand{\barepropS}[4]{\tikz[baseline=-2.5pt]{
\draw[Aactivity]   (-0.3,0) --(0,0)   node[at start, above] {$#1$}  node[at start, below] {$#2$};
\draw[Aactivity] (0.3,0) -- (0,0) node[at start,above] {$#3$}
node[at start, below] {$#4$};
}}

\newcommand{\fricpert}[2]{
\tikz[baseline=-2.5pt]{
\draw[Aactivity]  (0,0)--(180:0.3)  node[at end ,above]{$#1$};
\draw[Aactivity]  (0,0)--(0:0.3)  node[at end ,above]{$#2$};
\draw[Bactivity] (270:0.5) -- (0,0) ;
\draw[red,fill=red] ([xshift=-2pt,yshift=-2pt]0,-0.5) rectangle ++(4pt,4pt);
}}

\newcommand{\fricpertshort}[2]{
\tikz[baseline=-2.5pt]{
\draw[Aactivity]  (0,0)--(180:0.2)  node[at end ,above]{$#1$};
\draw[Aactivity]  (0,0)--(0:0.2)  node[at end ,above]{$#2$};
\draw[Bactivity] (270:0.5) -- (0,0) ;
\draw[red,fill=red] ([xshift=-2pt,yshift=-2pt]0,-0.5) rectangle ++(4pt,4pt);
}}

\newcommand{\fricpertshortshort}[2]{
\tikz[baseline=-2.5pt]{
\draw[Aactivity]  (0,0)--(180:0.2)  node[at end ,above]{$#1$};
\draw[Aactivity]  (0,0)--(0:0.1)  node[at end ,above]{$#2$};
\draw[Bactivity] (270:0.5) -- (0,0) ;
\draw[red,fill=red] ([xshift=-1pt,yshift=-2pt]0,-0.5) rectangle ++(4pt,4pt);
}}

\newcommand{\fricpertX}[4]{
\tikz[baseline=-2.5pt]{
\draw[Aactivity]  (0,0)--(180:0.3)  node[at end ,above]{$#1$} node[at end ,below]{$#3$};
\draw[Aactivity]  (0,0)--(0:0.3)  node[at end ,above]{$#2$} node[at end ,below]{$#4$};
\draw[Bactivity] (270:0.5) -- (0,0) ;
\draw[red,fill=red] ([xshift=-2pt,yshift=-2pt]0,-0.5) rectangle ++(4pt,4pt);
}}

\newcommand{\velopertXshort}[4]{
\tikz[baseline=-2.5pt]{
\draw[Aactivity]  (0,0)--(180:0.3)  node[at end ,above]{$#1$} node[at end ,below]{$#3$};
\draw[Aactivity]  (0,0)--(0:0.1)  node[at end ,above]{$#2$} node[at end ,below]{$#4$};
\draw[Baractivity]  (-0.2,-0.1)--(-0.2,0.1);
\draw[red,fill=red]  circle(1mm);
}}

\newcommand{\velopertX}[4]{
\tikz[baseline=-2.5pt]{
\draw[Aactivity]  (0,0)--(180:0.3)  node[at end ,above]{$#1$} node[at end ,below]{$#3$};
\draw[Aactivity]  (0,0)--(0:0.3)  node[at end ,above]{$#2$} node[at end ,below]{$#4$};
\draw[Baractivity]  (-0.2,-0.1)--(-0.2,0.1);
\draw[red,fill=red]  circle(1mm);
}}
\newcommand{\fricpertXshort}[4]{
\tikz[baseline=-2.5pt]{
\draw[Aactivity]  (0,0)--(180:0.3)  node[at end ,above]{$#1$} node[at end ,below]{$#3$};
\draw[Aactivity]  (0,0)--(0:0.1)  node[at end ,above]{$#2$} node[at end ,below]{$#4$};
\draw[Bactivity] (270:0.5) -- (0,0) ;
\draw[red,fill=red] ([xshift=-2pt,yshift=-2pt]0,-0.5) rectangle ++(4pt,4pt);
}}

\newcommand{
\HermiteSimple}[1]{H\! e_{#1}}

\begin{document}
\title{Field Theory of Active Brownian Particles with Dry Friction}

\author{Ziluo Zhang}
\affiliation{Wenzhou Institute, University of Chinese Academy of Sciences, Wenzhou, Zhejiang 325001, China}
\affiliation{Institute of Theoretical Physics, Chinese Academy of Sciences, Beijing 100190, China}

\author{Shurui Yuan}
\affiliation{Harbin Institute of Technology,  Harbin, 150001, China}
\affiliation{Wenzhou Institute, University of Chinese Academy of Sciences, Wenzhou, Zhejiang 325001, China}

\author{Shigeyuki Komura}
\thanks{Corresponding author \href{mailto:komura@wiucas.ac.cn}{komura@wiucas.ac.cn} }
\affiliation{Wenzhou Institute, University of Chinese Academy of Sciences, Wenzhou, Zhejiang 325001, China}
\affiliation{Oujiang Laboratory, Wenzhou, Zhejiang 325000, China}
\affiliation{Department of Chemistry, Graduate School of Science, Tokyo Metropolitan University, Tokyo 192-0397, Japan}

\begin{abstract}
We present a field theoretic approach to capture the motion of a particle with dry friction for one- and two-dimensional  diffusive particles, and further expand the framework for two-dimensional active Brownian particles. 
Starting with the Fokker-Planck equation and introducing the Hermite polynomials as the corresponding eigen-functions, we obtain the actions and propagators.
Using a perturbation expansion, we calculate the effective diffusion coefficient in the presence of both wet and dry frictions in a 
perturbative way via the Green-Kubo relation. 
We further compare the analytical result with the numerical simulation. Our result can be used to estimate the values of dry friction coefficient in experiments.
\end{abstract}

\maketitle

\section{Introduction}

Active matter~\cite{Cates:2012,Bechinger:2016,Marchetti:2013,Ramaswamy:2010} refers to particles or creatures that 
can absorb energy from the environment and change their mechanical state by using the energy.  The collective behaviors of animals
such as swarming~\cite{Mach:2007, Vollmer:2006}, flocks~\cite{Cavagna_2018,Reynolds:1987,Toner:1995,Toner:1998,Vicsek:1995},  
and bacteria~\cite{Berg:2004} can be described effectively by appropriate models. Furthermore, artificial active colloidal particles~\cite{Fily:2012, Huang:2024} are created experimentally to explore active systems. There are plenty of analytical works from the Stokes–Einstein–Sutherland relation of Brownian motion~\cite{Doi_textbook} to active systems~\cite{Tailleur:2008, RobertsPruessner:2022}. However, fitting analytical results with experimental ones is always a challenge. For example, the restricted confinement can significantly affect the diffusion coefficient \cite{Avni:2021}. Furthermore, the motion of particles can be influenced by dry friction (Coulomb friction) with the substrate if the size and weight of the particle are large enough. In this case, the buoyancy and thermal fluctuations can barely push the particle away from the substrate. Unlike the wet friction that is always proportional to the velocity, dry friction depends only on the velocity direction.

In the most of the previous studies, dry and wet frictions are not considered together. The particles experiencing dry friction are mostly unaffected by the thermal noise, since the particles are large enough. Moreover, the gravity acting on the  particles is usually  smaller than thermal noise, and the particles hardly reach the lower surface. Hence, dry friction of such particles can also be neglected.  
For mesoscopic particles, however,  both wet and dry frictions can not be neglected because the gravity drags the particles and their random motions take place at the bottom surface \cite{Xie_He_2019,Lin_He_2017}. A one-dimensional (1D) passive particle with dry friction has been discussed \cite{deGennes_2005, HAYAKAWA_2005, Touchette_2010}, and further two-dimensional (2D) problem has been also addressed \cite{Das_2017}. 

In this work, we are motivated to explore the motion of an active particle with dry friction. In particular, we obtain the effective diffusion coefficient of a 2D active Brownian particle (ABP) with both dry and wet frictions by using the Doi-Peliti field theory~\cite{Doi:1976,Peliti:1985,Cardy:2008}. A 2D ABP moves with a finite velocity, and its director undergoes rotational diffusion~\cite{Cates:2013, Kurzthaler:2016, Stenhammar:2014}. We first introduce the Langevin equations to describe the motion of an ABP, and later convert it to the corresponding Fokker-Planck equation. Since the effective diffusion coefficient can be calculated via the velocity-velocity correlation function by using the Green-Kubo relation \cite{Doi_textbook}, we only consider the particle velocity in this work.

 In Sec.~\ref{sec:math}, we introduce the mathematical basis of this work, such as the equation of motion and its 
 eigen-system to simplify the calculation. 
 In Sec.~\ref{sec:1d_diffusion}, we use the field-theoretic framework to capture the motion of a 1D diffusing particle with dry friction in a perturbative way.  We expand our framework to 2D space in Sec.~\ref{sec:2d_diffusion} by approximating dry friction to be isotropic.
By treating the self-propulsion of an ABP as a perturbative part, we obtain the corresponding effective diffusion coefficient in Sec.~\ref{sec:abp}. We also compare the 2D result with the simulation result.  A summary of our work and some discussion are given in Sec.~\ref{sec:discussion}.

\section{Model}
\label{sec:math}
In this section, we firstly show the mathematical basis of our work. Our main aim is to calculate the effective diffusion coefficient $\effeD$ as a function of dry friction coefficient  $\friction$. The diffusion coefficient $\effeD$ can be extracted from the  mean squared displacement in the long-time limit. In this work, we obtain the effective diffusion coefficient by the  Green-Kubo relation \cite{Doi_textbook},
\begin{equation}
\elabel{Green_Kubo}
    \effeD=\lim_{t\rightarrow \infty} \frac{\ave{\big(\xvec(t)
    -\xvec(0)\big)^2}}{2dt}=\frac{1}{d}\int_0^\infty \dint t\ave{\vvec(t)\cdot\vvec(0)}\ ,
\end{equation}
where the $\ave{\bullet}$ is the ensemble average and $d$ is the space dimension.

 We describe the motion of an ABP  with dry friction by the Langevin equations 
\begin{subequations}
    \begin{align}
        m\dot{\vvec}(t)&=-\Gamma\vvec(t)-F\frac{\vvec(t)}{||\vvec(t)||}+\Gamma\wvec_\theta(t)+\Xivec(t) &&\quad\text{with}\quad\ave{\Xivec(t)\Xivec^\text{T}(t')}=2 D \Gamma^2\ident_2 \delta(t-t')\quad \text{and}\quad\ave{\Xivec}=\nullvec\ ,\\
         \dot{\theta}(t)&=\zeta(t)
    &&\quad\text{with}\quad
    \ave{\zeta(t)\zeta(t')} = 2\rotDiffusion\delta(t-t')\quad \text{and}\quad\ave{\zeta(t)}=0 \ ,\\
    \dot{\xvec}&=\vvec\ ,
    \end{align}
\end{subequations}
where $||\vvec||$ is the norm of the vector $\vvec$, $\wvec_\theta(t)=w[\cos\theta,\sin\theta]^\text{T}$ is the self-propelled velocity of the particle which stays on a ring of 
 a radius of $w$, and $m$ is the mass of  the particle. Furthermore, $\Gamma\vvec(t)$ is the viscous force proportional to the velocity and $F \vvec(t)/||\vvec(t)||$ is dry friction force whose magnitude is a constant. And $\ident_2$ is a 2D unit matrix. Each component of  $\Xivec$ and also $\zeta$ are Gaussian white noise \cite{VanKampen:2007} with zero mean and variance $D\Gamma^2$ and $\rotDiffusion$, respectively, where  $D$ and $\rotDiffusion$ are the diffusion and rotational diffusion coefficients respectively. And each component of $\Xivec$ has zero correlation with $\zeta$. Finally, $\xvec$ is the position of  the particle, and we further define the dry friction term $\vvec/||\vvec||=\lim_{\epsilonvec\rightarrow\nullvec}\vvec/||\vvec+\epsilonvec||$ to avoid the zero denominator.

Introducing the rescaled friction coefficients by the mass $\gamma=\Gamma/m$ and $\friction=F/m$, we modify the Langevin equation as
 \begin{subequations}
 \elabel{Langevin_general}
    \begin{align}
        \dot{\vvec}(t)&=-\gamma\vvec(t)-\friction\frac{\vvec(t)}{||\vvec(t)||}+\gamma\wvec_\theta(t)+\xivec(t) &&\quad\text{with}\quad\ave{\xivec(t)\xivec^\text{T}(t')}=2 D \gamma^2\ident_2 \delta(t-t') \quad \text{and}\quad\ave{\xivec(t)}=\nullvec \  ,\\
         \dot{\theta}(t)&=\zeta(t)
    &&\quad\text{with}\quad
    \ave{\zeta(t)\zeta(t')} = 2\rotDiffusion\delta(t-t')\quad \text{and}\quad\ave{\zeta(t)}=0 \ ,
    \end{align}
\end{subequations}
whose corresponding Fokker-Planck equation is \cite{Risken:1989, Pavliotis:2014}
\begin{equation}
\elabel{FPE_1}
    \partial_t P(\vvec, \theta,t)=\operator P(\vvec, \theta,t) \quad \text{with}\quad \operator=D\gamma^2\nabla^2_\vvec+\gamma\nabla_\vvec \cdot\vvec+\friction \nabla_\vvec \cdot \frac{\vvec}{||\vvec||}-\wvec_\theta\cdot\nabla_\vvec+\rotDiffusion \partial_\theta^2\ .
\end{equation}
Since the diffusion coefficient can be extracted from the velocity-velocity correlation function, we drop the positional variable $\xvec$ in the Fokker-Planck equation, and we  only consider the particle in the velocity space. The effective diffusion coefficient of an isolated 2D ABP in free space is well known \cite{Cates:2015, Zhang:2024}, in our case, it presents the zero dry friction limit $\friction=0$. We write it below for the later comparison
\begin{equation}
\elabel{effe_D_iso_ABP}
    D_0=D+\frac{w^2}{2\rotDiffusion} \ .
\end{equation}

The following functions are used later to simplify the calculation,
\begin{subequations}
\elabel{hermite_def}
    \begin{align}
        u_n(v)&=\exp{-\frac{v^2}{2\effeV^2}}\Hermite{n}{\frac{v}{\effeV}}\ ,\\
        \utilde_n(v)&=\frac{1}{\sqrt{2\pi}n!}\Hermite{n}{\frac{v}{\effeV}} \ , 
    \end{align}
\end{subequations}
where $\Hermite{n}{x}$ is the $n$-th order of the \emph{probabilist's Hermite polynomials} and $\effeV^2=D\gamma$ \cite{abramowitzStegun, Garcia-MillanPruessner:2021}. 
Our definition of $\Hermite{n}{x}$ corresponds to $2^{-\frac{n}{2}}$\texttt{HermiteH[$n,x/\sqrt{2}$]} \cite{Mathematica:10.0.2.0}.

The orthogonality relation between $u_n(v)$ and $\utilde(v)$ is 
\begin{equation}
\elabel{ortho_relation}
    \int \dint v u_n(v) \utilde_m(v)=\effeV \delta_{n,m}\ ,
\end{equation}
where $\delta_{n,m}$ is the  Kronecker delta, and $u_n(v)$ are the eigenfunctions of the operator
\begin{equation}
    \effeV\partial_v^2 u_n(v)+\partial_v \big[v u_n(v)\big]=-n u_n(v)\ .
\end{equation}


The velocity-velocity correlation function with the initial velocity $\vvec_0$, direction $\theta_0$ at time $t_0$ can be calculated by 
\begin{align}
\elabel{velocity-velocity_correlation}
\ave{\vvec(t)\cdot\vvec(t')}=&\ave{\vvec\cdot\vvec'}=\int \dTWOint{v} \dTWOint{v'}\int_0^{2\pi}\dint{\theta}\dint{\theta'} \vvec  \Propagator\big(\vvec,\theta,t|\vvec',\theta',t'\big) \vvec'  \Propagator\big(\vvec',\theta',t'|\vvec_0,\theta_0,t_0\big) \ ,
\end{align}
where we use $\vvec(t)=\vvec$ and $\vvec(t')=\vvec'$ to simplify the notations and $\Propagator(\vvec,\theta,t|\vvec',\theta',t')$ is the probability density of finding a particle at velocity $\vvec$ with the self-propulsion direction $\theta$ at time $t$, when the given initial state is  at velocity $\vvec'$ with the propulsion director $\theta'$ at time $t'$.

Since we choose the stationary state as the initial state, the most right propagator $\Propagator\big(\vvec',\theta',t'|\vvec_0,\theta_0,t_0\big)= \Propagator(\vvec',\theta') $ does not depend on the initial condition.
By using the orthogonality of the Hermite polynomials \cite{RGMTHESIS}, the above integral can be simplified further, and the result will be presented in the following sections.

\section{Diffusive particle in 1D space}
\label{sec:1d_diffusion}

We start with the simplest case, i.e., a diffusive particle with 
 dry friction in 1D space. 
The corresponding Fokker-Planck equation is
\begin{equation}
\elabel{one_d_FPE}
    \partial_t P(v,t)=D\gamma^2 \partial^2_v P(v,t)+\gamma \partial_v v P(v,t)+\mu \partial_v \frac{v}{|v|} P(v,t) \ ,
\end{equation}
where in 1D space, the unit vector $\vvec/||\vvec||$ becomes the sign function $v/|v|$ where $|v|$ is the absolute value of $v$.
We further introduce the notations
\begin{subequations}
    \begin{align}
        \operator_0&=D\gamma^2\partial_v^2+\gamma\partial_v v \ ,\\
        \operator_\friction &=\mu\partial_v \frac{v}{|v|} \ ,
    \end{align}
\end{subequations}
similar to the 2D case, we define $v/|v|=\lim_{\epsilon\rightarrow 0}v/|v+\epsilon|$ to avoid the zero denominator.
\subsubsection{Action}
The corresponding bilinear action and perturbative action are
\begin{subequations}
    \begin{align}
    \elabel{one_d_action_split}
    \action&=\action_0+\actionPert \ ,\\
    \elabel{one_d_action_0}
    \action_0(\phi,\phitilde)&=\int\dint{t}\int\dint{v}\phitilde(v,t) \big(-\partial_t +\operator_0-\mass\big)\phi(v,t) \ ,\\
\elabel{one_d_action_friction}\actionPert(\phi,\phitilde)&=\int\dint{t}\int\dint{v}\phitilde(v,t) \operator_\friction\phi(v,t) \ .
    \end{align}
\end{subequations}
where $\mass$ is the death rate to maintain the causality, and will be taken to zero after the inverse Fourier transform.  We have introduced the annihilation field $\phi$ and the Doi-shifted creation field $\phitilde$  as \cite{Doi:1976, Cardy:2008}
\begin{subequations}
\elabel{one_d_fields}
    \begin{align}
        \phi(v,t)&=\frac{1}{\effeV}\int\dbar{\omega}\exp{-\imag\omega t}\sum_{n=0}^\infty u_n(v)\phi_n(\omega) \ ,\\
        \phitilde(v,t)&=\frac{1}{\effeV}\int\dbar{\omega}\exp{-\imag\omega t}\sum_{n=0}^\infty \utilde_n(v)\phitilde_n(\omega) \ .
    \end{align}
\end{subequations}
Here $\effeV^2=D\gamma$, $\dbar \omega=\dint \omega/(2\pi)$ and further $\deltabar(\omega)=2\pi\delta(\omega)$, while 
$u$ and $\utilde$ are consistent with \Eref{hermite_def}. 
Any expectation value can be calculated perturbatively by a path integral \cite{PruessnerGarcia-Millan:2022}
\begin{equation}
	\ave{\bullet} = \int\Dint{\phi}\Dint{\phitilde} \bullet\ \exp{\action[\phitilde,\phi]} \ =\ave{\bullet\ \exp{\actionPert[\phitilde,\phi]}}_0 \quad\text{with}\quad \ave{\bullet}_0=\int\Dint{\phi}\Dint{\phitilde}\bullet\ \exp{\action_0[\phitilde,\phi]}\ ,
\end{equation}
where we split the action via \Eref{one_d_action_split}. Expanding the exponential with respect to the perturbative part  
$\actionPert$, we obtain
\begin{equation}\elabel{perturbative_action_use}
    \ave{\bullet} = \int\Dint{\phi}\Dint{\phitilde} \exp{\action_0[\phitilde,\phi]}\bullet\sum_{n=0}^\infty\frac{\actionPert^n}{n!} 
    = \ave{\bullet\sum_{n=0}^\infty\frac{\actionPert^n}{n!}}_0 \ .
\end{equation}
The probability density is the full propagator 
\begin{equation}
\Propagator(\vvec,\theta,t|\vvec_0,\theta_0,t_0)=\ave{\phi(\vvec,\theta,t)\phitilde(\vvec_0,\theta_0,t_0)} \ .
\end{equation}

By plugging the fields \Eref{one_d_fields} into the action \Eref{one_d_action_0}, we obtain
\begin{align}
\elabel{1d_action_0}
    \action_0&=-\frac{1}{\effeV^2}\int\dbar{\omega}\int\dbar{\omega}' \sum_{n,m=0}^\infty \phitilde_m(\omega')\big(-\imag\omega+\gamma n+\mass\big)\phi_n(\omega) \deltabar(\omega+\omega')\effeV\delta_{n,m}\ ,
\end{align}
where $\effeV\delta_{n,m}$ comes from the orthogonality relation \Eref{ortho_relation} between $u_n$ and $u_m$. The bare propagator can be read off as
\begin{equation}
     \ave{\phi_n(\omega)\phitilde_m(\omega')}_0=
    \frac{\effeV \delta_{n,m}\deltabar(\omega+\omega')}{-\imag\omega +\gamma n +\mass}=\effeV\delta_{n,m}\deltabar(\omega+\omega')\propga{n}{\omega}\corresponds\bareproplong{n,\omega}{m,\omega'} \ .
\end{equation}
Similarly, we substitute the fields in \Eref{one_d_fields} into the perturbative action in \Eref{one_d_action_friction},
\begin{subequations}
\elabel{1d_action_p}
\begin{align}
    \actionPert&=\sum_{n,m}\frac{\mu}{\effeV^2}\int\dbar{\omega}  \dbar{\omega'}\phitilde_m(\omega')\phi_n(\omega)\int \dint v \Tilde{u}_m(v) \partial_v \big[\frac{v}{|v|} u_n(v)\big]\deltabar(\omega+\omega')\\
    &=\sum_{n,m}\frac{\mu}{\effeV^2}\int\dbar{\omega}  \phitilde_m(-\omega)\phi_n(\omega)\int \dint v \bigg(2\delta(v)\utilde_m(v) u_n(v)+\frac{v}{|v|}\utilde_m(v)\partial_v  u_n(v)\bigg)\ ,
\end{align}
\end{subequations}
where  $\delta(v)$ comes from the derivative of the sign function
\begin{equation}
    \partial_v\left(\frac{v}{|v|}\right)=2\delta(v)\ .
\end{equation}
We diagrammatically write the perturbative part of the action as 
\begin{equation}
   \frac{\mu}{\effeV^2}\int \dint v \bigg(2\delta(v)\utilde_m(v) u_n(v)+\frac{v}{|v|}\utilde_m(v)\partial_v  u_n(v)\bigg)=\Lambda^{m,n} \corresponds \fricpert{m}{n} \ .
\end{equation}
By using the following properties of the Hermite polynomials \cite{abramowitzStegun},
\begin{subequations}
\begin{align}
    \Hermite{n}{v}\Hermite{m}{v}&=\sum_{k=0}^{\min{(n,m)}} \frac{n! m!}{(n-k)!(m-k)! k!}\Hermite{n+m-2k}{v} \ ,\\
    \partial_v \big(\Hermite{n}{v}\exp{-\frac{v^2}{2}}\big)&=-\Hermite{n+1}{v}\exp{-\frac{v^2}{2}}\ ,
\end{align}
\end{subequations}
the analytic form of the above vertex becomes
\begin{equation}
    \fricpert{m}{n} = \frac{\friction}{\effeV^2}\frac{2}{\sqrt{2\pi}m!}\bigg(\Hermite{n}{0}\Hermite{m}{0}-\sum_{k=0}^{\min{(n+1,m)}} \frac{(n+1)! m!}{(n+1-k)!(m-k)! k!} \Hermite{n+m-2k}{0}\bigg) \ ,
 \end{equation}
 where $\Hermite{n}{0}$ is  the ``Hermite zero'', which is probabilist's Hermite polynomials evaluated at zero with respect to the $n$-th order.
It is trivial to see that
\begin{equation}
\elabel{1D_0n_Lambda}
    \Lambda^{0,n}=0 \ \ \ \text{for} \ \ n\in \mathbb{Z}^+_0 \ ,
\end{equation}
which indicates that a propagator does not have an outgoing index $0$ unless it is a bare one. Moreover, since the ``Hermite zero'' of arbitrary odd order is zero, $\Hermite{2n+1}{0}=0$, we find
    \begin{align}
        \Lambda^{2m+1,2n}=\Lambda^{2m,2n+1}=0 \ \ \ \text{for} \ \ n,m\in \mathbb{Z}^+_0\ ,
    \end{align}
which shows the outgoing and incoming indices should have the same parity. Otherwise, the corresponding combinations become zero.

In the following, we only consider the perturbative vertices with the limited indices $n,m=0,1,2$, which are
\begin{subequations}
\elabel{1d_vertices}
\begin{align}
        \Lambda^{2,0}&=\Lambda^{1,1}=\Lambda^{2,2}=-\sqrt{\frac{2}{
        \pi}}\frac{\mu}{\effeV^2}\ ,\\
\Lambda^{0,0}&=\Lambda^{0,1}=\Lambda^{1,0}=\Lambda^{0,2}=\Lambda^{1,2}=\Lambda^{2,1}=0 \ .
    \end{align}
\end{subequations}

\subsubsection{Propagator}
By using the bare propagator and perturbative vertices, the full propagator can be written in a perturbative way as
\begin{subequations}
\elabel{1d_full_prop}
    \begin{align}
        \ave{\phi_n(\omega)\phitilde_m(\omega')}&\corresponds \bareprop{n,\omega}{m,\omega'}+\bareprop{n,\omega}{}\!\!\!\!\fricpert{}{}\!\!\!\!\bareprop{}{m,\omega'}+\bareprop{n,\omega}{}\!\!\!\!\fricpert{}{}\!\!\!\!\bareprop{}{}\!\!\!\!\fricpert{}{}\!\!\!\!\bareprop{}{m,\omega'}+\dots\\
        &=\deltabar(\omega+\omega')\big\{\effeV\delta_{n,m}\propga{n}{\omega}+\effeV^2 \propga{n}{\omega}\Lambda^{n,m}\propga{m}{\omega}+\dots \big\} \ .
    \end{align}
\end{subequations}
The full propagators is therefore
\begin{equation}
\elabel{one_d_full_prop}
\ave{\phi_n(\omega)\phitilde_m(\omega')}=\deltabar(\omega+\omega')\sum_{j=0}^\infty E_j(n,m,\omega) \ ,
\end{equation}
and the corresponding $E_j$ are given by
\begin{subequations}
\begin{align}
    E_0(n,m,\omega)&=\effeV \delta_{n,m}\propga{n}{\omega}\\
    E_1(n,m,\omega)&=\effeV^2 \propga{n}{\omega}\Lambda^{n,m}\propga{m}{\omega} \\
    E_2(n,m,\omega)&=\effeV^3 \propga{n}{\omega}\sum_{q=0}^\infty\Lambda^{n,q}\propga{q}{\omega}\Lambda^{q,m}\propga{m}{\omega}    =\effeV\propga{n}{\omega}\sum_{q=0}^\infty \Lambda^{n,q}E_1(q,m,\omega) \ .
\end{align}
\end{subequations}
We further introduce the recurrence relation between $E_j$ and $E_{j+1}$
\begin{align}
\elabel{one_d_recurrence_time}
    E_{j+1}(n,m,\omega)&=\effeV\propga{n}{\omega}\sum_{q=0}^\infty \Lambda^{n,q}E_j(q,m,\omega)
\end{align}
The probability density of velocity $v$ at time $t$ with a given initial state $v'$ at time $t'$ is 
\begin{subequations}
\elabel{one_d_density}
\begin{align}
    \Propagator(v,t|v',t')&=\frac{1}{\effeV^2}\sum_{n,m=0}^\infty u_n(v) \utilde_m(v')
    \lim_{\mass\downarrow 0}
    \int \dbar{\omega}\dbar{\omega'}\exp{-\imag \omega t}\exp{-\imag \omega' t'}\ave{\phi_n(\omega)\phitilde_m(\omega')}\\
    &= \frac{1}{\effeV^2}\sum_{n,m=0}^\infty u_n(v) \utilde_m(v') \lim_{\mass\downarrow 0}\int \dbar{\omega}\exp{-\imag \omega (t-t')}\sum_j E_j(n,m,\omega)\ .
\end{align}
\end{subequations}

\subsubsection{Stationary-state correlation function}
By taking the limit $t_0\rightarrow -\infty$, we obtain the stationary density
\begin{equation}
    \Propagator(v)=\lim_{t_0\rightarrow -\infty}\ave{\phi(v,t)\phitilde(v_0,t_0)} \ .
\end{equation}
Previous work shows that taking the stationary limit $t_0\rightarrow-\infty$ and the zero death rate limit $\mass\downarrow 0$ replaces the incoming  index by $\delta_{m,0}$ \cite{Zhang:2024}, which physically indicates that the steady-state is independent of the initial condition. Therefore, at stationarity, \Eref{one_d_recurrence_time} is 
\begin{align}
    E_j(n)&=\lim_{\mass\downarrow 0}\lim_{t_0\rightarrow-\infty} \int \dbar{\omega} \exp{-\imag \omega(t-t_0)}  E_j(n,m,\omega) \\
    &= \bareprop{n}{}\!\!\!\!\!\!\!\!\underbrace{\fricpert{}{}\!\!\!\!\bareprop\!\!\!\!\!\!\!\!\fricpert{}{}\!\!\!\!\dots\bareprop\!\!\!\!\!\!\!\!\fricpertshortshort{}{}}_{j \ \text{vertices}} \nonumber \ ,
\end{align}
where we have dropped the arguments $m,\omega$, since the observable no longer depends on  time and initial state. The form of the stationary density is 
\begin{equation}
\elabel{one_d_sta_density}
    \Propagator(v)=\frac{1}{\effeV^2}\sum_{n=0}^\infty u_n(v) \utilde_0\sum_{j=0}^\infty E_j(n) \ .
\end{equation}
Since $\utilde_0(v)$ does not depend on $v$, we use the notation $\utilde_0\corresponds \utilde_0(v)$ to indicate the independence of the initial state.
We list the first two orders and the recurrence relation of $E_j$ as follows
\begin{subequations}
\elabel{one_d_recurs}
    \begin{align}
        E_0(n)&= \effeV \delta_{n,0} \ , \\
        E_1(n)&= \frac{\Lambda^{n,0} \effeV^2}{n \gamma}\  , \\
        E_{j+1}(n)&=  \frac{\Lambda^{n,q} \effeV}{n \gamma} \sum_{q=1}E_j(q) \elabel{one_d_recur} \ .
    \end{align}
\end{subequations}
Since $\Lambda^{0,n}$ is always zero in \Eref{1D_0n_Lambda} which vanishes all the propagators with outgoing index $n=0$ before taking stationary limit,  we define $\lim_{n\rightarrow 0} \Lambda^{n,m}/n=0$ for arbitrary $m$ and perform the summation from $q=1$ to avoid the zero denominators in \Eref{one_d_recurs}. 

Our aim is to use the Green-Kubo relation \Eref{Green_Kubo} to calculate the effective diffusion coefficient.  Now, we perform the integral of  the velocity-velocity  correlation function in the present field theory framework
\begin{align}
        \ave{v(t)v(t')}&=\int \dint{v} \int \dint{v'} v \Propagator(v,t|v',t') v' \Propagator(v')    \ ,
\end{align}
where we simplify the notation $v=v(t)$ and $v'=v(t')$. By using Eqs.~\eref{velocity-velocity_correlation}, \eref{one_d_density}, \eref{one_d_sta_density} and the orthogonality of the Hermite polynomials \cite{RGMTHESIS}, 
the correlation function becomes
\begin{equation}
\elabel{eq_to_corr}
    \ave{v(t)v(t')}= \int\dbar\omega \exp{-\imag \omega(t-t')} \ave{\phi_1(\omega)\phitilde_1(-\omega)} \bigg(2\ave{\phi_2 \phitilde_0}+\ave{\phi_0  \phitilde}\bigg) \ .
\end{equation}
We first calculate the time-independent observable in the brackets as
\begin{subequations}
\elabel{1D_time_independent_observable}
    \begin{align}
        \ave{\phi_0\phitilde_0}&=\effeV \elabel{one_d_field_00}\ , \\
        \elabel{one_d_field_20} \ave{\phi_2\phitilde_0}&=\sum_{j=0} E_j(2)=-\effeV \frac{\rescalefric}{2+\rescalefric}+\mathcal{O}(\rescalefric^2) \ ,
    \end{align}
\end{subequations}
where in \Eref{one_d_field_20} we only consider the finite number of perturbative vertices up to the second order, and $\mathcal{O}(\rescalefric^2)$ is used to indicate that we do not consider all the contributions higher than the second order of $\rescalefric$. The mathematical details are shown in Appendix~\ref{app:observables_pure}. We introduce a dimensionless parameter 
\begin{equation}
\elabel{1d_dimensionless_para}
\rescalefric=\sqrt{\frac{2}{\pi}}\frac{\mu}{\effeV \gamma}    \ .
\end{equation}
to simplify the notation.
Then we have
\begin{equation}
    2\ave{\phi_2 (0) \phitilde_0(0)}+\ave{\phi_0 (0) \phitilde(0)}=\effeV\bigg(1-\frac{\rescalefric}{1+\frac{\rescalefric}{2}}\bigg) +\mathcal{O}(\rescalefric^2)\ . \elabel{1d_corr_time_indepen} 
\end{equation}

To calculate the time-dependent observable, we only consider the perturbative part with both incoming and outgoing indices fixed to unity, 
\begin{subequations}
    \begin{align}
        &\int\dbar\omega \exp{-\imag \omega(t-t')} \ave{\phi_1(\omega)\phitilde_1(-\omega)} \nonumber\\
        &=\int\dbar\omega \exp{-\imag \omega(t-t')}  \bareprop{1,\omega}{1,\omega'}+\bareprop{1,\omega}{}\!\!\!\!\fricpert{}{}\!\!\!\!\bareprop{}{1,\omega'}+\bareprop{1,\omega}{}\!\!\!\!\fricpert{}{}\!\!\!\!\!\!\!\!\bareprop{1}{}\!\!\!\!\fricpert{}{}\!\!\!\!\bareprop{}{1,\omega'}+\dots\\
        &=\int\dbar\omega \exp{-\imag \omega(t-t')} \effeV \propga{1}{\omega}  \sum_{\ell=0}^\infty \big(\effeV \propga{1}{\omega}  \Lambda^{1,1}\big)^\ell+\mathcal{O}(\rescalefric^2)\\
        &=\int\dbar\omega \exp{-\imag \omega(t-t')}  \frac{\effeV \propga{1}{\omega}}{1-\effeV\propga{1}{\omega}\Lambda^{1,1} }+\mathcal{O}(\rescalefric^2)\\
        &=
\int \dbar{\omega} \  \exp{-\imag \omega(t-t')} \frac{\effeV}{-\imag \omega+\gamma+\sqrt{\frac{2}{\pi}}\frac{\mu}{\effeV}}+\mathcal{O}(\rescalefric^2)=\Theta(t-t') \effeV \exp{-(\gamma +\gamma\rescalefric)(t-t')}  +\mathcal{O}(
\rescalefric^2)   \ , \elabel{1d_corr_time_depen}
\end{align}
\end{subequations}
where $\Theta$ is the Heaviside step function, since the integral converges only for $t\geq t'$.
Then, the time-correlation function is obtained by substituting Eqs.~\eref{1d_corr_time_indepen} and \eref{1d_corr_time_depen} into \Eref{eq_to_corr}, 
\begin{equation}
\elabel{1d_corr_func}
    \ave{v(t)v(t')}=\Theta(t-t')\effeV^2\exp{-\gamma(1 +\rescalefric)(t-t')}  \bigg(1-\frac{\rescalefric}{1+\frac{\rescalefric}{2}} \bigg)+ \mathcal{O}(
\rescalefric^2)  \ .
\end{equation}
By using the Green-Kubo relation in \Eref{Green_Kubo} and recalling $\effeV^2=D\gamma$, the effective diffusion coefficient is 
\begin{equation}
\elabel{1d_effe_D}
     \effeD= \int_{t'}^\infty \dint t\ave{v(t)v(t')}=\frac{D}{1+\rescalefric} \bigg(1-\frac{\rescalefric}{1+\frac{\rescalefric}{2}} \bigg)+\mathcal{O}(
\rescalefric^2)   \ . 
\end{equation}
If there is no dry friction, $\beta\rightarrow 0$, we recover the normal translational diffusion coefficient $\effeD|_{\rescalefric\rightarrow 0}=D$, as it should. By expanding the fractions, we obtain the first order  correction with respect to the  parameter $\rescalefric$ as
\begin{equation}
\elabel{1d_first_order_correction}
    \effeD=D(1-2\rescalefric)+\mathcal{O}(\rescalefric^2)\ .
\end{equation}

\section{Diffusive particle in 2D space}
\label{sec:2d_diffusion}

In 2D space, the corresponding  Fokker-Planck equation of a diffusive particle with dry friction is
\begin{equation}
    \partial_t P(\vvec,t)=D\gamma^2 \nabla^2_\vvec P(\vvec,t)+\gamma \nabla_\vvec \cdot \big[\vvec P(\vvec,t)\big]+\mu \nabla_\vvec \cdot\big[\frac{\vvec}{||\vvec||} P(\vvec,t)\big] \ .
\end{equation}
Similar to the 1D case, we split the operator into two parts,
\begin{subequations}
    \begin{align}
        \operator_0&=D\gamma^2 \nabla^2_\vvec +\gamma \nabla_\vvec \cdot \vvec \ , \\
        \operator_\mu&=\mu \nabla_\vvec \cdot\frac{\vvec}{||\vvec||} \ .
    \end{align}
\end{subequations}
Applying the isotropic property of the first operator $\operator_0$, one can simplify the calculation into 1D space with an extra prefactor $2$. However, because of the anisotropy of dry friction, there is no such  a way to simplify the operator $\operator_\mu$. Hence,  we use an approximation of the ``sign'' function of the velocity $\vvec$
\begin{equation}
\elabel{2D_fric_approx}
    \frac{\vvec}{||\vvec||}=\begin{bmatrix}
        \frac{v_1}{\sqrt{v_1^2+v_2^2}}\\
        \\
         \frac{v_2}{\sqrt{v_2^2+v_2^2}}
    \end{bmatrix}=\begin{bmatrix}
        \frac{v_1}{|v_1|\sqrt{1+\frac{v_2^2}{v_1^2}}}\\
        \\
         \frac{v_2}{|v_2|\sqrt{1+\frac{v_1^2}{v_2^2}}}
    \end{bmatrix}\approx \exp{-\frac{1}{4}}\begin{bmatrix}
        \frac{v_1}{|v_1|}\\
        \\
         \frac{v_2}{|v_2|}
    \end{bmatrix} \ ,
\end{equation}
where we use the following Hermite expansion of the square root in the denominator, and the prefactor $\exp{-1/4}$ is the projection of the dry friction term on the $
\HermiteSimple{0}$,
\begin{equation}
    \frac{1}{\sqrt{1+z^2}}=\exp{-\frac{1}{4}}
    \sum_{k=0}^\infty {\frac{(-1)^k}{2^k(2k)!}}\Hermite{2k}{\sqrt{2} \theta} \ ,
\end{equation}
with $\theta=\arctan(z)$.
By considering only the zeroth order Hermite polynomial and dropping all the higher order terms, we find  the approximation in \Eref{2D_fric_approx}. The operator $\operator_\mu$ is rewritten as 
\begin{equation}
\elabel{friction_approx}
    \operator_\mu\approx \operator'_\mu=\mu \exp{-\frac{1}{4}}\bigg(\partial_{v_1} \frac{v_1}{|v_1|}+\partial_{v_2} \frac{v_2}{|v_2|}\bigg) \ ,
\end{equation}
which is now isotropic.
Then, we can simplify this 2D problem to a 1D problem whose corresponding Fokker-Planck equation is
\begin{equation}
\elabel{2d_iso_FPE}
    \partial_t P^{(2)}(v,t)=D \gamma^2 \partial_v^2 P^{(2)}(v,t)+\gamma\partial_v v P^{(2)}(v,t) +\exp{-\frac{1}{4}}\mu\partial_v \frac{v}{|v|} P^{(2)}(v,t)\ ,
\end{equation} 
where we have introduced an upper index $^{(2)}$ to indicate that the above Fokker-Planck equation is obtained from the 2D equation. The corresponding velocity-velocity correlation function is
\begin{equation}
    \ave{\vvec(t)\cdot\vvec(t')}=2\ave{v(t)v(t')}\ ,
\end{equation}
where the prefactor $2$ in the RHS is the dimension factor. 

The only difference between the Fokker-Planck equation in \Eref{2d_iso_FPE} and 1D case in \Eref{one_d_FPE} is an extra prefactor 
$\exp{-1/4}$ in the friction term. Therefore, by following the same steps presented in 1D case in  Sec.~\ref{sec:1d_diffusion}, we immediately obtain the velocity-velocity correlation and the effective diffusion coefficient by introducing  the modified dimensionless parameter as
\begin{equation}
\elabel{2d_dimensionless_para}
\rescalefricTwoD=\exp{-\frac{1}{4}}\sqrt{\frac{2}{\pi}}\frac{\mu}{\effeV \gamma}    \ ,
\end{equation}
and they are
\begin{subequations}
\begin{align}
        \ave{\vvec(t)\cdot\vvec(t')}&=2D\gamma\exp{-\gamma(1+\rescalefricTwoD)(t-t')}\bigg(1-\frac{\rescalefricTwoD}{1+\frac{\rescalefricTwoD}{2}}  \bigg)+\mathcal{O}(\rescalefricTwoD^2) \ , \\
     \effeD^{(2)}&= \frac{D}{1+\rescalefricTwoD}\bigg(1-\frac{\rescalefricTwoD}{1+\frac{\rescalefricTwoD}{2}}  \bigg)+\mathcal{O}(\rescalefricTwoD^2)\ .\elabel{2D_diffusion_effe_D}
\end{align}
\end{subequations}
Here we use an upper index $ ^{(2)}$ to distinguish the effective diffusion coefficient in 2D from the 1D result in \Eref{1d_effe_D}, where the parameter $\rescalefricTwoD$ in the 2D problem is changed from the parameter  $\rescalefric$ in the 1D case.

We compare our result in \Eref{2D_diffusion_effe_D} with the numerical simulation. In this work, only finite types of the perturbation vertices are considered, the friction operator is approximated by \Eref{friction_approx}, and the result is a perturbation calculation using the first three orders of the Hermite polynomials. We compare the effective diffusion coefficient with  the numerical simulation and our field theory approach in Fig.~\ref{Fig.main}(a),  showing a good agreement between them even $\rescalefricTwoD$ is large. Additionally, the fourth order correction of the time-independent observable $\ave{\phi_2\phitilde_0}$ is also calculated in Appendix~\ref{app:observables_pure}.

\section{Active Brownian Particle}
\label{sec:abp}

ABPs are particles that move with constant speed $|\wvec|=w$ but whose director $\theta$ diffuses with the rotational diffusion coefficient $\rotDiffusion$. The corresponding Fokker-Planck equation  of a 2D ABP with dry friction is
\begin{equation}
\elabel{ABP_fpe}
    \partial_t P(\vvec, \theta,t)=\operator P(\vvec, \theta,t) \quad \text{with}\quad \operator=D\gamma^2\nabla_\vvec+\gamma\nabla_\vvec \cdot\vvec+\friction \nabla \cdot \frac{\vvec}{||\vvec||}-\gamma\wvec_\theta\cdot\nabla+\rotDiffusion \partial_\theta^2\ ,
\end{equation}
where $\wvec_\theta=w[\cos\theta,\sin\theta]^\text{T}$.

Similar to the approach presented in Sec.~\ref{sec:2d_diffusion}, we first use the approximation of the friction term in Eqs.~\eref{2D_fric_approx} and \eref{friction_approx}, to approximate the anisotropic operator by an isotropic one. Second, we reduce the problem to 1D and choose the component of $\vvec$ on the $x$-axis of the Cartesian plane. The Fokker-Planck equation is then
\begin{equation}
    \partial_t P^{(2)}(v, \theta,t)=\big(\operator_0 +\operator_\mu+\operator_w\big) P^{(2)}(v,\theta,t) 
\end{equation}
with
\begin{subequations}
\elabel{ABP_operator}
    \begin{align}
        \operator_0&= D\gamma^2\partial_v^2+\gamma \partial_v v+\rotDiffusion\partial_\theta^2 \ ,\\
        \operator_\mu&\approx\operator'_\mu=\mu\exp{-\frac{1}{4}} \partial_v \frac{v}{|v|} \ ,\\
        \operator_w&= -\gamma w\cos{\theta} \partial_v \ .
    \end{align}
\end{subequations}

\subsection{Action}
Accordingly, the bilinear action and perturbative actions are
\begin{subequations}
    \begin{align}
        \action_0(\phi,\phitilde)&=\int\dint{t} \int\dint{v} \int_0^{2\pi}\dint\theta \phitilde(v,\theta,t)\big(-\partial_t+\operator_0-\mass\big)\phi(v,\theta,t) \ ,\\
        \action_\mu(\phi,\phitilde)&=\int\dint{t} \int\dint{v} \int_0^{2\pi}\dint\theta \phitilde(v,\theta,t)\operator'_\mu\phi(v,\theta,t)\ ,\\
        \action_w(\phi,\phitilde)&=\int\dint{t} \int\dint{v} \int_0^{2\pi}\dint\theta \phitilde(v,\theta,t)\operator_w\phi(v,\theta,t)\ .
    \end{align}
\end{subequations}
Since there is a new variable $\theta$, the fields in this case are introduced as
\begin{subequations}
\elabel{ABP_fields}
    \begin{align}
        \phi(v,\theta,t)&=\frac{1}{\effeV}\sum_{n=0}^\infty\sum_{\alpha=-\infty}^\infty\int \dbar{\omega}\exp{-\imag \omega t} u_n(v)\exp{-\imag \alpha\theta}\phi_{n,\alpha}(\omega) \ ,\\
         \phitilde(v,\theta,t)&=\frac{1}{2\pi\effeV}\sum_{n=0}^\infty\sum_{\alpha=-\infty}^\infty\int \dbar{\omega}\exp{-\imag \omega t} \utilde_n(v)\exp{\imag \alpha\theta}\phitilde_{n,\alpha}(\omega) \ .
    \end{align}
\end{subequations}
Here, we use Roman letters for the velocity indices and Greek letters for the director indices.
Similar to the 1D case, the corresponding bilinear action and the $\mu$-dependent perturbative action are
\begin{subequations}
\begin{align}
    \action_0&=-\frac{1}{2\pi\effeV^2}\sum_{n,m=0}^\infty\sum_{\alpha,\alpha'=-\infty}^\infty  \int\dbar{\omega}\dbar{\omega'}
\phitilde_{m,\alpha'}(\omega')\big(-\imag\omega+\gamma n+ \rotDiffusion \alpha^2+\mass\big)\phi_{n,\alpha}(\omega)\deltabar(\omega+\omega')\effeV\delta_{n,m}2\pi\delta_{\alpha,\alpha'}\ ,\\
\action_\mu&=\frac{\mu\exp{-1/4}}{2\pi\effeV^2}\sum_{n,m=0}^\infty\sum_{\alpha,\alpha'=-\infty}^\infty  \int\dbar{\omega}  \phitilde_{m,\alpha'}(-\omega)\phi_{n,\alpha}(\omega)\int \dint v \bigg(2\delta(v)\utilde_m(v) u_n(v)+\frac{v}{|v|}\utilde_m(v)\partial_v  u_n(v)\bigg)2\pi\delta_{\alpha,\alpha'}\ ,
\end{align}
\end{subequations}
and the self-motility dependent action $\action_w$ is
\begin{subequations}
\begin{align}
    \action_w&=-\frac{\gamma w}{2\pi\effeV^2}\sum_{n,m=0}^\infty\sum_{\alpha,\alpha'=-\infty}^\infty  \int\dbar{\omega}\phitilde_{m,\alpha'}(-\omega)\phi_{n,\alpha}(\omega) \int \dint{v} \utilde_m(v) \partial_v u_n(v) \int_0^{2\pi}\dint{\theta}\exp{-\imag(\alpha-\alpha') \theta}\cos{\theta}\\
    &=\frac{\gamma w}{2\pi\effeV^2}\sum_{n,m=0}^\infty\sum_{\alpha,\alpha'=-\infty}^\infty  \int\dbar{\omega}\phitilde_{m,\alpha'}(-\omega)\phi_{n,\alpha}(\omega)  \delta_{n,m-1}2\pi\frac{\delta_{\alpha+1,\alpha'}+\delta_{\alpha-1,\alpha'}}{2} \ .
\end{align}
\end{subequations}

There are two different types of perturbative vertices. We diagrammatically write the friction-dependent and self-propulsion dependent vertices in the following way
\begin{subequations}
\elabel{abp_vertices}
    \begin{align}
         \frac{\mu \exp{-\frac{1}{4}}}{\effeV^2}\int \dint v \bigg(2\delta(v)\utilde_m(v) u_n(v)+\frac{v}{|v|}\utilde_m(v)\partial_v  u_n(v)\bigg)\delta_{\alpha,\alpha'}&\corresponds\Lambda_{\alpha',\alpha}^{m,n}\corresponds\fricpertX{m}{n}{\alpha'}{\alpha} \ ,\\ 
         \gamma w\frac{\delta_{\alpha+1,\alpha'}+\delta_{\alpha-1,\alpha'}}{2\effeV^2} \delta_{n,m-1}&\corresponds\Upsilon_{\alpha',\alpha}^{m,n}\corresponds\velopertX{m}{n}{\alpha'}{\alpha}\ ,
    \end{align}
\end{subequations}
and the  bare propagator is
\begin{equation}
\elabel{abp_bare_prop}
    \frac{\effeV \delta_{n,m}\delta_{\alpha,\alpha'}\deltabar(\omega+\omega')}{-\imag\omega +\gamma n+\rotDiffusion \alpha^2+\mass}=\effeV\delta_{n,m}\delta_{\alpha,\alpha'}\deltabar(\omega+\omega')\propga{n,\alpha}{\omega}\corresponds\barepropX{n,\omega}{\alpha}{m,\omega'}{\alpha'}\ .
\end{equation}

\subsection{Propagator}
The full propagator is the summation of all the possible combination of the bare propagator in \Eref{abp_bare_prop} and the perturbative vertices in \Eref{abp_vertices}
\begin{subequations}
    \begin{align}
        &\ave{\phi_{n,\alpha}(\omega)\phitilde_{m,\alpha'}(\omega')}\corresponds \barepropX{n,\omega}{\alpha}{m,\omega'}{\alpha'}+\barepropS{n,\omega}{\alpha}{}{}\!\!\!\!\fricpertX{}{}{}{}\!\!\!\!\barepropS{}{}{m,\omega'}{\alpha'}+\barepropS{n,\omega}{\alpha}{}{}\!\!\!\!\velopertX{}{}{}{}\!\!\!\!\barepropS{}{}{m,\omega'}{\alpha'}+\nonumber\\
        &\barepropS{n,\omega}{\alpha}{}{}\!\!\!\!\velopertX{}{}{}{}\!\!\!\!\!\!\!\!\!\!\!\barepropS{
        \sum_q}{\sum_{\nu}}{}{}\!\!\!\!\velopertX{}{}{}{}\!\!\!\!\barepropS{}{}{m,\omega'}{\alpha'}
        +\barepropS{n,\omega}{\alpha}{}{}\!\!\!\!\fricpertX{}{}{}{}\!\!\!\!\!\!\!\!\!\!\!\barepropS{\sum_q}{\sum_{\nu}}{}{}\!\!\!\!\velopertX{}{}{}{}\!\!\!\!\barepropS{}{}{m,\omega'}{\alpha'}
    +\barepropS{n,\omega}{\alpha}{}{}\!\!\!\!\velopertX{}{}{}{}\!\!\!\!\!\!\!\!\!\!\!\barepropS{\sum_q}{\sum_{\nu}}{}{}\!\!\!\!\fricpertX{}{}{}{}\!\!\!\!\barepropS{}{}{m,\omega'}{\alpha'}
    +\barepropS{n,\omega}{\alpha}{}{}\!\!\!\!\fricpertX{}{}{}{}\!\!\!\!\!\!\!\!\!\!\!\barepropS{\sum_q}{\sum_{\nu}}{}{}\!\!\!\!\fricpertX{}{}{}{}\!\!\!\!\barepropS{}{}{m,\omega'}{\alpha'}
    +\dots
        \\&=\deltabar(\omega+\omega')\bigg\{\effeV\delta_{n,m}\delta_{\alpha,
        \alpha'}\propga{n,\alpha}{\omega}+\effeV^2\propga{n,\alpha}{\omega}\big(\Upsilon_{n,m}^{\alpha,\alpha'}+\Lambda_{n,m}^{\alpha,\alpha'}\big)\propga{m,\alpha'}{\omega} \nonumber\\
        \ \ \ \ \ &+\effeV^3\propga{n,\alpha}{\omega}\sum_{q=0}^\infty\sum_{\nu=-\infty}^\infty\big(\Upsilon_{n,q}^{\alpha,\nu}+\Lambda_{n,q}^{\alpha,\nu}\big)\propga{q,\nu}{\omega}\big(\Upsilon_{q,m}^{\nu,\alpha'}+\Lambda_{q,m}^{\nu,\alpha'}\big)\propga{m,\alpha'}{\omega}+\dots\bigg\} \elabel{ABP_first_third_order}\ .
    \end{align}
\end{subequations}
Similar to \Eref{one_d_full_prop}, we write the full propagator as, 
\begin{equation}
\ave{\phi_{n,\alpha}(\omega)\phitilde_{m,\alpha'}(\omega')}=\deltabar(\omega+\omega')\sum_{j=0}^\infty E_j(n,m,\alpha,\alpha',\omega) \ ,  
\end{equation}
where the terms shown in \Eref{ABP_first_third_order} are exactly the first three orders of $E_j(n,m,\alpha,\alpha',\omega)$. We list the recurrence relation as
\begin{align}
    E_{j+1}(n,m,\alpha,\alpha',\omega)=\effeV \propga{n,\alpha}{\omega}\sum_{q=0}^\infty\sum_{\nu=-\infty}^\infty\big(\Upsilon_{n,q}^{\alpha,\nu}+\Lambda_{n,q}^{\alpha,\nu}\big)E_{j}(q,m,\nu,\alpha',\omega) \ .
\end{align}
We therefore obtain the probability density of the velocity at $v$ with director $\theta$ at time $t$ with the corresponding initial state $(v',\theta',t')$ as
\begin{subequations}
\begin{align}
    \Propagator(v,\theta,t|v',\theta',t')&=\frac{1}{2\pi\effeV^2}\sum_{n,m=0}^\infty\sum_{\alpha,\alpha'=-\infty}^\infty u_n(v)\utilde_m(v')\exp{-\imag \alpha\theta} \exp{\imag\alpha'\theta'} \lim_{\mass\downarrow 0}\int\dbar\omega\dbar\omega' \deltabar(\omega+\omega')\ave{\phi_{n,\alpha}(\omega)\phitilde_{m,\alpha'}(\omega')}\\
    &=\frac{1}{2\pi\effeV^2}\sum_{n,m=0}^\infty\sum_{\alpha,\alpha'=-\infty}^\infty u_n(v)\utilde_m(v')\exp{-\imag \alpha\theta} \exp{\imag\alpha'\theta'} \lim_{\mass\downarrow 0}\int\dbar\omega \exp{-\imag \omega(t-t')}\sum_{j=0}E_j(n,m,\alpha,\alpha',\omega)\ .
    \elabel{ABP_full_prop_sum}
\end{align}
\end{subequations}

\subsection{Velocity-velocity correlation function}
By substituting \Eref{ABP_full_prop_sum} into \Eref{velocity-velocity_correlation}, and using the orthogonality of the Hermite polynomials and the exponential terms, we have
\begin{equation}
    \ave{v(t)v(t')}=\ave{\phi_{1,0}(t)\phitilde_{1,0}(t')} \bigg(2\ave{\phi_{2,0}\phitilde_{0,0}}+\ave{\phi_{0,0}\phitilde_{0,0}}\bigg)+
    \sum_{\nu=-\infty}^\infty\ave{\phi_{1,0}(t) \phitilde_{0,\nu}(t')}\ave{\phi_{1,\nu}\phitilde_{0,0}} \ ,
\end{equation}
where the second term comes from the integral over $\theta'$. Since we only consider the first three orders of the speed indices $n,m=0,1,2$ of the perturbation vertices in \Eref{abp_vertices}, only $\nu=\pm1$ in the summation are concerned  in the second term.  We calculate the time-independent observable first by the inverse Fourier transform and take the limits $t_0\rightarrow -\infty$ and $r\downarrow0$
\begin{equation}
\ave{\phi_{n,\alpha}\phitilde_{m,\alpha'}}=\lim_{r\downarrow0}\lim_{t_0\rightarrow -\infty}\int\dbar\omega \exp{-\imag \omega (t-t_0)}\ave{\phi_{n,\alpha}(\omega)\phitilde_{m,\alpha'}(-\omega)}\ .
\end{equation}
We obtain
\begin{subequations}
\elabel{ABP_time_independent_observable}
    \begin{align}
    \elabel{ABP_time_independent_observable_1}
        \ave{\phi_{0,0}\phitilde_{0,0}}&=\effeV \ ,\\
        \elabel{ABP_time_independent_observable_2}
        \ave{\phi_{2,0}\phitilde_{0,0}}&=-\effeV \frac{\rescalefricTwoD}{2+\rescalefricTwoD}+\frac{w^2\gamma}{4\effeV(\gamma+\rotDiffusion+\rescalefricTwoD \gamma)(1+\frac{\rescalefricTwoD}{2})}+\mathcal{O}(\rescalefricTwoD^2) \ ,\\ \elabel{ABP_time_independent_observable_3}
\ave{\phi_{1,1}\phitilde_{0,0}}=\ave{\phi_{1,-1}\phitilde_{0,0}}&=\frac{w\gamma}{2(\gamma+\rotDiffusion+\rescalefricTwoD\gamma)}+\mathcal{O}(\rescalefricTwoD^2)\ ,
    \end{align}
\end{subequations}
where the details are presented in Appendix~\ref{app:observables_abp}.

\begin{figure}[H]
\centering  
\includegraphics[width=17cm]{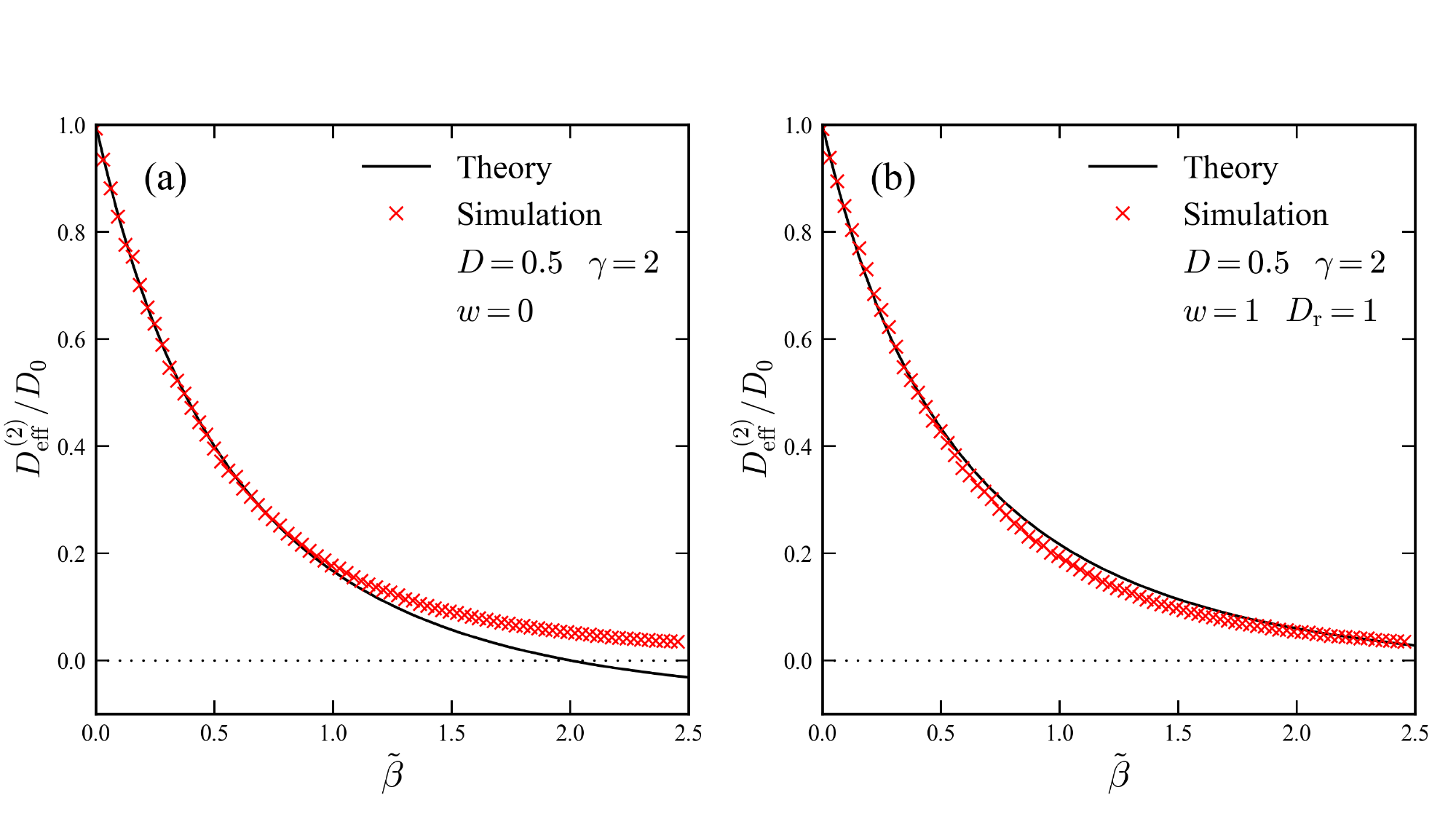}
\caption{Comparison between the analytical result and numerical simulation. The horizontal axis is the dimensionless parameter $\rescalefricTwoD$ in \Eref{2d_dimensionless_para}, and the vertical axis is the ratio between the effective diffusion coefficient $\effeD^{(2)}$ in Eqs.~\eref{2D_diffusion_effe_D}, \eref{abp_effD} and the diffusion coefficient $D_0$ in the absence of dry friction in \Eref{effeD_abp_no_fric}.   The numerical result of $\effeD^{(2)}$ is obtained from the Langevin equation in \Eref{Langevin_general}  via the Euler–Maruyama method \cite{kloeden:2011}, where the deatils are explained in Appendix~ \ref{app:Numerical}. We have used the following parameters: number of the particles $N=20000$, total time $T=400$, and time step $\Delta t=0.05$. The initial state of the simulation is $\xvec(-50)=\nullvec$ and $\theta(-50)=0$, where we have spent $T_0=50$ to wait for the system to reach the stationary state and then started to average the MSD. }
\label{Fig.main}
\end{figure}

For the time-dependent correlation function, we only consider the lowest perturbation vertices, similar to Sec.~\ref{sec:1d_diffusion}.   The first one is
\begin{subequations}
    \begin{align}
         & \ave{\phi_{1,0}(t)\phitilde_{1,0 }(t')} =\int\dbar\omega \exp{-\imag \omega(t-t')} \ave{\phi_{1,0}(\omega)\phitilde_{1,0}(-\omega)}\\
        &=\int\dbar\omega \exp{-\imag \omega(t-t')}  \barepropX{1,\omega}{0}{1,-\omega}{0}+\barepropX{1,\omega}{0}{}{}\!\!\!\!\fricpert{}{}\!\!\!\!\bareprop{}{1,-\omega}+\bareprop{1,\omega}{}\!\!\!\!\fricpert{}{}\!\!\!\!\!\!\!\!\bareprop{1}{}\!\!\!\!\fricpert{}{}\!\!\!\!\barepropX{}{}{1,-\omega}{0}+\dots\\
        &=\int\dbar\omega \exp{-\imag \omega(t-t')} \effeV \propga{1,0}{\omega}  \sum_{\ell=0}^\infty \big(\effeV \propga{1,0}{\omega}  \Lambda^{1,1}_{0,0}\big)^\ell+\mathcal{O}(\rescalefricTwoD^2)\\
&=\int \dbar{\omega} \  \exp{-\imag \omega(t-t')} \frac{\effeV}{-\imag \omega+\gamma+\exp{-\frac{1}{4}}\sqrt{\frac{2}{\pi}}\frac{\mu}{\effeV}}=\Theta(t-t') \effeV \exp{-(\gamma +\gamma\rescalefricTwoD)(t-t')} +\mathcal{O}(\rescalefricTwoD^2)\ ,
    \end{align}
\end{subequations}
where $\rescalefricTwoD$ is the corresponding dimensionless parameter introduced in \Eref{2d_dimensionless_para}. The second term is
\begin{subequations}
    \begin{align}
& \ave{\phi_{1,0}(t)\phitilde_{0,1 }(t')}=\ave{\phi_{1,0}(t)\phitilde_{0,-1 }(t')} =\int\dbar\omega \exp{-\imag \omega(t-t')} \ave{\phi_{1,0}(\omega)\phitilde_{0,1 }(-\omega)} \\
&=\int\dbar\omega \exp{-\imag \omega(t-t')} \barepropX{1,\omega}{0}{}{}\!\!\!\!\velopertX{}{}\!\!\!\!\!\!\!\!\barepropX{}{}{0,-\omega}{1}+\barepropX{1,\omega}{0}{}{}\!\!\!\!\fricpertX{}{1}{}{0}\!\!\!\!\!\!\!\!\barepropX{}{}{}{}\!\!\!\!\velopertX{}{}{}{}\!\!\!\!\barepropX{}{}{0,-\omega}{1}\nonumber\\ &+\barepropX{1,\omega}{0}{}{}\!\!\!\!\fricpertX{}{1}{}{0}\!\!\!\!\!\!\!\!\barepropX{}{}{}{}\!\!\!\!\fricpertX{}{1}{}{0}\!\!\!\!\!\!\!\!\barepropX{}{}{}{}\!\!\!\!\velopertX{}{}{}{}\!\!\!\!\barepropX{}{}{0,-\omega}{1}\dots\\
&=\int\dbar\omega \exp{-\imag \omega(t-t')} \effeV^2 \propga{1,0}{\omega}\Upsilon_{0,1}^{1,0}\propga{0,1}{\omega} \sum_{\ell=0}^\infty \big(\effeV \propga{1,0}{\omega}\Lambda_{0,0}^{1,1}\big)^\ell+\mathcal{O}(\rescalefricTwoD^2)\\
&= \Theta(t-t') \frac{w\gamma}{2(\gamma+\rescalefricTwoD \gamma-\rotDiffusion)} \bigg(\exp{-\rotDiffusion (t-t')}-\exp{-\gamma(1+\rescalefricTwoD) (t-t')}\bigg)+\mathcal{O}(\rescalefricTwoD^2) \ .
    \end{align}
\end{subequations}
Then, the velocity-velocity correlation function is 
\begin{align}
\elabel{abp_velo_corr}
\ave{v(t)v(t')}=&\Theta(t-t') \bigg(\effeV^2 \exp{-\gamma(1+\rescalefricTwoD)  (t-t')}\big(1-\frac{\rescalefricTwoD}{1+\frac{\rescalefricTwoD}{2}}\big)\nonumber\\
&+\frac{\gamma w^2 \Expp{-\gamma(1+\rescalefricTwoD) (t-t')}}{2(\gamma+\rotDiffusion+\gamma\rescalefricTwoD)(1+\frac{\rescalefricTwoD}{2})}+\frac{\Expp{-\rotDiffusion (t-t')}-\Expp{-\gamma(1+\rescalefricTwoD) (t-t')}}{\gamma+\rescalefricTwoD \gamma-\rotDiffusion} \times\frac{w^2\gamma^2}{2(\gamma+\rotDiffusion+\rescalefricTwoD\gamma ) }\bigg) +\mathcal{O}(\rescalefricTwoD^2)\ .
\end{align}
By applying the Green-Kubo relation in \Eref{Green_Kubo}, we obtain the effective diffusion coefficient for an ABP with dry friction as
\begin{equation}
\elabel{abp_effD}
    \effeD^{(2)}=\frac{D}{1+\rescalefricTwoD}\big(1-\frac{\rescalefricTwoD}{1+\frac{\rescalefricTwoD}{2}}\big)+\frac{w^2 }{2(\gamma+\rotDiffusion+\gamma\rescalefricTwoD)(1+\rescalefricTwoD)(1+\frac{\rescalefricTwoD}{2})}+\frac{w^2 \gamma}{2\rotDiffusion(\gamma+\rotDiffusion+\gamma\rescalefricTwoD)(1+\rescalefricTwoD)} +\mathcal{O}(\rescalefricTwoD^2)\ .
\end{equation}
When there is no friction, $\rescalefricTwoD=0$, we recover the effective diffusion for an isolated ABP in \Eref{effe_D_iso_ABP} \cite{Scholz:2018,Patel:2023,Cates:2015,Zhang:2024},
\begin{equation}
\elabel{effeD_abp_no_fric}
    \effeD^{(2)}\big|_{\rescalefricTwoD\rightarrow 0}=D+\frac{w^2}{2\rotDiffusion}= D_0 \ .
\end{equation}
Similar to \Eref{1d_first_order_correction}, by expanding the fractions, we obtain the first order correction of the effective diffusion coefficient for an ABP with dry friction as
\begin{equation}
\elabel{abp_effD_first_order}
    \effeD^{(2)}=D(1-2\rescalefricTwoD)+\frac{w^2}{2\rotDiffusion}\left(1-\frac{3\rotDiffusion+4\gamma}{2(\rotDiffusion+\gamma)}\rescalefricTwoD\right)+\mathcal{O}(\rescalefricTwoD^2)\ .
\end{equation}
We further compare the analytical result in \Eref{abp_effD} with the numerical simulation in Fig.~\ref{Fig.main}(b). Even with the existence of the self-propulsion, the agreement between the  field  theoretic approach and the simulation is still good.

\section{Summary and discussion}
\label{sec:discussion}

In this paper, we first discussed the 1D diffusing particle with dry friction. From the velocity-corresponding Fokker-Planck equation in \Eref{one_d_FPE}, we determined the bilinear and perturbative actions for a diffusive particle with dry friction in Eqs.~\eref{1d_action_0} and \eref{1d_action_p}, respectively. The full propagator was derived in \Eref{1d_full_prop}. Considering the first three orders of the Hermite polynomials, we only count three types of perturbative vertices with non-zero contributions  \Eref{1d_vertices}. Therefore, the summation in the propagator is reduced to a single term, and the geometric sum is used to obtain the velocity-velocity correlation function 
 in \Eref{1d_corr_func}.  Equation \eref{1d_effe_D}
was further obtained by using the Green-Kubo relation \Eref{Green_Kubo}.

Then, we have extended this framework to the 2D space. 
For wet friction, we can always reduce the higher dimensional problem to a 1D problem because the system is isotropic.
For dry friction, however, such a treatment is no longer possible because it is anisotropic. 
We then introduced a Hermite expansion of dry friction, and transformed dry friction into an isotropic operator in \Eref{2D_fric_approx} by considering the leading order.
With a prefactor difference, we obtained the 2D effective diffusion coefficient \Eref{2D_diffusion_effe_D} from 
the 1D  result in \Eref{1d_effe_D}.

Studying the ABP problem is more complicated since the self-propulsion of the particle provides another type of perturbation vertex. 
By using the same treatment as in the 1D case, we only considered the limited Hermite order 
($n\leq2 $). 
Then, we applied the geometric sums to calculate the velocity-velocity correlation function in \Eref{abp_velo_corr} 
and further obtained the effective diffusion coefficient in \Eref{abp_effD}.  
Since we have neglected higher orders of the Hermite expansion for both fields and dry friction operator, 
the diffusion coefficient is a perturbative result.
However, the analytical result in \Eref{abp_effD} recovers the numerical simulation very well.

Dry friction force $F$ in the experiment can be estimated by using \Eref{abp_effD} or the first order expansion in \Eref{abp_effD_first_order}.
Substituting back $\Gamma=m \gamma$ and $F=m \friction$, and using the Stokes–Einstein–Sutherland relations $\Gamma=6\pi \eta a$ and $D=k_{\rm B} T/\Gamma$ \cite{Doi_textbook}, where $\eta$ is the fluid viscosity, $a$ is the radius of the particle, $k_{\text{B}}$ is the  Boltzmann constant and $T$ is the temperature, we 
rewrite the dimensionless parameter $\rescalefricTwoD$  as
\begin{equation}
    \rescalefricTwoD= \exp{-\frac{1}{4}}F \sqrt{\frac{2 m}{\pi D \Gamma^3} }= \exp{-\frac{1}{4}}\frac{F}{6\pi\eta a} \sqrt{\frac{2 m}{\pi k_{\text{B}} T} }\ .
\end{equation}
Considering a colloidal particle with a mass $m \approx 10^{-15} \text{kg}$ and a radius $a \approx 10^{-6} \text{m}$ in water with the room temperature, and assuming $F \approx 0.1 \times mg$, where
$g$ is the gravitational constant, 
we estimate  $\rescalefricTwoD\approx 5\times 10^{-4}$.
If the radius is increased to $a\approx 5 \times10^{-6} \text{m}$, the corresponding parameter becomes $\rescalefricTwoD\approx 0.14$, which leads to a substantial decrease of the effective diffusion coefficient.


The present work provides a basis for the characterization of the motion of active matter in velocity space via field theory, especially for mesoscopic particles whose diffusion and mass can not be neglected. Because of the wet friction $\gamma$, the calculation is simplified by using  Hermite expansion of the fields \cite{Garcia-MillanPruessner:2021} rather than the Fourier transform \cite{Zhang:2024}. This work also provides a basis for ABPs with harmonic interactions. With the aid of the previous work on field theoretic approach of interacting diffusive particles \cite{Zhang:2023, PruessnerGarcia-Millan:2022}, we are currently working on ABPs with non-reciprocal harmonic interactions.

\section*{Acknowledgements}
S.K.\ acknowledges the support by National Natural Science Foundation of China (Nos.\ 12274098 and 
12250710127) and the startup grant of Wenzhou Institute, University of Chinese Academy of Sciences 
(No.\ WIUCASQD2021041).
This work was supported by the JSPS Core-to-Core Program ``Advanced core-to-core network for the physics of self-organizing active matter" (JPJSCCA20230002).

\bibliography{articles}
\bibliographystyle{unsrt}

\appendix

\section{Observables}
\label{app:observables}

In the following, we show the calculation of the observable $\ave{\bullet}$ in detail.

\subsection{Purely diffusive case}
\label{app:observables_pure}

We first show \Eref{1D_time_independent_observable}. By using \Eref{1D_0n_Lambda}, we immediately obtain \Eref{one_d_field_00} since the bare propagator is the only contribution. For the observable in the LHS of \Eref{one_d_field_20}, we only consider the limited perturbation vertices listed in \Eref{1d_vertices}. Diagrammatically, it is
\begin{subequations}
\elabel{one_d_phi20}
    \begin{align}
        \ave{\phi_2\phitilde_0}&=\bareprop{2}{}\!\!\!\!\fricpertshortshort{}{0}+\bareprop{2}{}\!\!\!\!\fricpertshort{}{} \!\!\!\!\!\!\!\!\bareprop{2}{}\!\!\!\!\fricpertshortshort{}{0} +\bareprop{2}{}\!\!\!\!\fricpertshort{}{} \!\!\!\!\!\!\!\!\bareprop{2}{}\!\!\!\!\fricpertshort{}{} \!\!\!\!\!\!\!\!\bareprop{2}{}\!\!\!\!\fricpertshortshort{}{0} +\dots\\&=\bareprop{2}{}\!\!\!\!\fricpertshortshort{}{0} \sum_{k=0}^\infty \big(\bareprop{2}{}\!\!\!\!\fricpertshortshort{}{2}\big)^k+\mathcal{O}(\rescalefric^2)\\
        &=\bareprop{2}{}\!\!\!\!\fricpertshortshort{}{0} \frac{1}{1-\bareprop{2}{}\!\!\!\!\fricpertshortshort{}{2}}+\mathcal{O}(\rescalefric^2)\\
        &=\effeV \frac{\Lambda^{2,0}\effeV}{2\gamma}\frac{1}{1-\frac{\Lambda^{2,2}\effeV}{2\gamma}}+\mathcal{O}(\rescalefric^2)\ .
    \end{align}
\end{subequations}
We further include higher order terms up to the fourth order, and the above observable becomes 
\begin{subequations}
    \begin{align}
         \ave{\phi_2\phitilde_0}&=\bareprop{2}{}\!\!\!\!\fricpertshortshort{}{0} \sum_{k=0}^\infty \big(\bareprop{2}{}\!\!\!\!\fricpertshortshort{}{2}\big)^k \bigg(1+\sum_{\ell=1}^\infty \big(\bareprop{2}{}\!\!\!\!\fricpertshort{}{} \!\!\!\!\!\!\!\!\bareprop{4}{}\!\!\!\!\fricpertshortshort{}{2}\big)^\ell \sum_{j=0}^\infty\big(\bareprop{4}{}\!\!\!\!\fricpertshortshort{}{4}\big)^j \bigg) \nonumber\\\elabel{20_higher_order_2}
         &+\bareprop{2}{}\!\!\!\!\fricpertshort{}{} \!\!\!\!\!\!\!\!\bareprop{4}{}\!\!\!\!\fricpertshortshort{}{0}\sum_{k=0}^\infty \big(\bareprop{2}{}\!\!\!\!\fricpertshortshort{}{2}\big)^k  \sum_{\ell=0}^\infty \big(\bareprop{2}{}\!\!\!\!\fricpertshort{}{} \!\!\!\!\!\!\!\!\bareprop{4}{}\!\!\!\!\fricpertshortshort{}{2}\big)^\ell \sum_{j=0}^\infty\big(\bareprop{4}{}\!\!\!\!\fricpertshortshort{}{4}\big)^j\\
         &=\effeV \frac{\Lambda^{2,0}\effeV}{2\gamma}\frac{1}{1-\frac{\Lambda^{2,2}\effeV}{2\gamma}}\bigg(1+\frac{\frac{\Lambda^{2,4}\Lambda^{4,2}\effeV^2}{8\gamma^2}}{1-\frac{\Lambda^{2,4}\Lambda^{4,2}\effeV^2}{8\gamma^2}}\frac{1}{1-\frac{\Lambda^{4,4}\effeV}{4\gamma}}\bigg) +\effeV \frac{\Lambda^{2,4}\Lambda^{4,0}\effeV^2}{8\gamma^2} \frac{1}{1-\frac{\Lambda^{2,2}\effeV}{2\gamma}}\frac{1}{1-\frac{\Lambda^{2,4}\Lambda^{4,2}\effeV^2}{8\gamma^2}}\frac{1}{1-\frac{\Lambda^{4,4}\effeV}{4\gamma}} \elabel{20_higher_order_3}+\mathcal{O}(\rescalefric^2)\ ,
    \end{align}
\end{subequations}
where the first line of \Eref{20_higher_order_2} indicates that the propagator starts with a vertex $\fricpertshort{2}{0}$ from most left, and the first summation shows that there can be arbitrary number of the propagator $\bareprop{2}{}\!\!\!\!\fricpertshortshort{}{2}$ connecting to the left of the propagator. The terms in the bracket indicates that, if there is at least one $\bareprop{2}{}\!\!\!\!\fricpertshort{}{} \!\!\!\!\!\!\!\!\bareprop{4}{}\!\!\!\!\fricpertshortshort{}{2}$ connected, the propagator $\bareprop{4}{}\!\!\!\!\fricpertshortshort{}{4}$ can also appear as many as possible.  Similarly, in the second line of \Eref{20_higher_order_2}, the propagator $\bareprop{2}{}\!\!\!\!\fricpertshort{}{} \!\!\!\!\!\!\!\!\bareprop{4}{}\!\!\!\!\fricpertshortshort{}{0}$ is the leading order, and there can be arbitrary number of the three types of the propagators appearing in the summations. We perform the geometric sum and obtain \Eref{20_higher_order_3}.

Since the term $\Lambda \effeV/\gamma$ is proportional to the dimensionless parameters $\rescalefric$ and $\rescalefricTwoD$ introduced in \Eref{1d_dimensionless_para} and \Eref{2d_dimensionless_para} for 1D and 2D space, respectively, we obtain the effective diffusion coefficient with the fourth order correction as
\begin{subequations}
    \begin{align}
    \effeD&=\frac{D}{1+\rescalefric}\bigg[1-\frac{\rescalefric}{1+\frac{\rescalefric}{2}}\big(1-\frac{\frac{\rescalefric^2}{16}}{1+\frac{\rescalefric^2}{16}}\frac{1}{1+\frac{3\rescalefric}{8}}\big)-\frac{\rescalefric^2}{24}\frac{1}{1+\frac{\rescalefric}{2}}\frac{1}{1+\frac{\rescalefric^2}{16}}\frac{1}{1+\frac{3\rescalefric}{8}}\bigg]+\mathcal{O}(\rescalefric^2)\ , \ \ \text{for} \ \ 1\text{D}\\
    \elabel{higher_order_effeD}
        \effeD^{(2)}&=\frac{D}{1+\rescalefricTwoD}\bigg[1-\frac{\rescalefricTwoD}{1+\frac{\rescalefricTwoD}{2}}\big(1-\frac{\frac{\rescalefricTwoD^2}{16}}{1+\frac{\rescalefricTwoD^2}{16}}\frac{1}{1+\frac{3\rescalefricTwoD}{8}}\big)-\frac{\rescalefricTwoD^2}{24}\frac{1}{1+\frac{\rescalefricTwoD}{2}}\frac{1}{1+\frac{\rescalefricTwoD^2}{16}}\frac{1}{1+\frac{3\rescalefricTwoD}{8}}\bigg]+\mathcal{O}(\rescalefricTwoD^2)\ , \ \ \text{for} \ \ 2\text{D} .
    \end{align}
\end{subequations}
    
\subsection{ABP case}
\label{app:observables_abp}

In the following, we obtain \Eref{ABP_time_independent_observable}. Similar to the pure diffusive case, $\ave{\phi_{0,0}\phitilde_{0,0}}$ in \Eref{ABP_time_independent_observable_1} is a bare propagator without any perturbative vertices. We write the LHS of \Eref{ABP_time_independent_observable_2} diagrammatically
\begin{subequations}
\elabel{ABP_phi20}
    \begin{align}
        \ave{\phi_{2,0}\phitilde_{0,0}}&=\barepropX{2}{0}{}{}\!\!\!\!\!\!\fricpertXshort{}{0}{}{0}
        +\barepropX{2}{0}{}{}\!\!\!\!\!\!\fricpertXshort{}{2}{}{0}\!\!\!\!\!\!\!\!\barepropX{}{}{}{}\!\!\!\!\!\!\fricpertXshort{}{0}{}{0}+\barepropX{2}{0}{}{}\!\!\!\!\!\!\fricpertXshort{}{2}{}{0}\!\!\!\!\!\!\!\!\barepropX{}{}{}{}\!\!\!\!\!\!\fricpertXshort{}{2}{}{0}\!\!\!\!\!\!\!\!\barepropX{}{}{}{}\!\!\!\!\!\!\fricpertXshort{}{0}{}{0}+\dots\\
        &+ \elabel{ABP_phi20_2}
        \barepropX{2}{0}{}{}\!\!\!\!\!\!\velopertXshort{}{1}{}{\pm 1}\!\!\!\!\!\!\!\!\barepropX{}{}{}{}\!\!\!\!\!\!\velopertXshort{}{0}{}{0}+\barepropX{2}{0}{}{}\!\!\!\!\!\!\fricpertXshort{}{2}{}{0}\!\!\!\!\!\!\!\barepropX{}{}{}{}\!\!\!\!\!\!\velopertXshort{}{1}{}{\pm 1}\!\!\!\!\!\!\!\!\barepropX{}{}{}{}\!\!\!\!\!\!\velopertXshort{}{0}{}{0}
        \\
        &+\barepropX{2}{0}{}{}\!\!\!\!\!\!\velopertXshort{}{1}{}{\pm 1}\!\!\!\!\!\!\!\!\barepropX{}{}{}{}\!\!\!\!\!\!\fricpertXshort{}{1}{}{\pm1}\!\!\!\!\!\!\!\!\!\!\!\!\barepropX{}{}{}{}\!\!\!\!\!\!\!\!\!\!\!\velopertXshort{}{0}{}{0}\dots 
    \end{align}
\end{subequations}
where the first line is exactly \Eref{one_d_phi20}, which is a combination of the bare propagator and the friction dependent perturbation only. The second and third lines  includes the self-propulsion perturbation part, where the first term in \Eref{ABP_phi20_2} is the leading order. The rest terms show that arbitrary number of $\barepropX{1}{\pm1}{}{}\!\!\!\!\!\!\fricpertXshort{}{1}{}{\pm1}$ and $\barepropX{2}{0}{}{}\!\!\!\!\!\!\fricpertXshort{}{2}{}{0}$  can be inserted into the diagram in a suitable position.
Therefore, we write \Eref{ABP_phi20} 
\begin{subequations}
\elabel{ABP_phi20_final}
    \begin{align}
         \ave{\phi_{2,0}\phitilde_{0,0}}&=\bareprop{2}{}\!\!\!\!\fricpertshortshort{}{0} \sum_{k=0}^\infty \big(\bareprop{2}{}\!\!\!\!\fricpertshortshort{}{2}\big)^k\nonumber\\
         &+ 2\times\barepropX{2}{0}{}{}\!\!\!\!\!\!\velopertXshort{}{1}{}{\pm 1}\!\!\!\!\!\!\!\!\barepropX{}{}{}{}\!\!\!\!\!\!\velopertXshort{}{0}{}{0} \sum_{k=0}^\infty \big(\barepropX{2}{0}{}{}\!\!\!\!\!\!\fricpertXshort{}{2}{}{0}\big)^k \sum_{\ell=0}^\infty \big(\barepropX{1}{1}{}{}\!\!\!\!\!\!\fricpertXshort{}{1}{}{1}\big)^\ell +\dots \\
         &=\effeV \frac{\Lambda^{2,0}\effeV}{2\gamma}\frac{1}{1-\frac{\Lambda^{2,2}\effeV}{2\gamma}}+2 \effeV\frac{\Upsilon_{0,1}^{2,1}\Upsilon_{1,0}^{1,0}\effeV^2}{2\gamma (\gamma+\rotDiffusion)} \frac{1}{1-\frac{\Lambda_{0,0}^{2,2}\effeV}{2\gamma}}\frac{1}{1-\frac{\Lambda_{1,1}^{1,1}\effeV}{\gamma+\rotDiffusion}}+\mathcal{O}(\rescalefricTwoD^2)\ , \elabel{ABP_phi20_final_2}
    \end{align}
\end{subequations}
where the prefactor $2$ is a symmetry factor that comes from $\pm1$.

Similary, for the observable in \Eref{ABP_time_independent_observable_3}, we  write it by the diagrams as
\begin{subequations}
    \begin{align}
    \ave{\phi_{1,1}\phitilde_{0,0}}&\approx\barepropX{1}{1}{}{}\!\!\!\!\!\!\velopertXshort{}{0}{}{0}+\barepropX{1}{1}{}{}\!\!\!\!\!\!\fricpertXshort{}{1}{}{1}\!\!\!\!\!\!\!\!\barepropX{}{}{}{}\!\!\!\!\!\!\velopertXshort{}{0}{}{0}+\barepropX{1}{1}{}{}\!\!\!\!\!\!\fricpertXshort{}{1}{}{1}\!\!\!\!\!\!\!\!\barepropX{ }{ }{}{}\!\!\!\!\!\!\fricpertXshort{}{1}{}{1}\!\!\!\!\!\!\!\!\barepropX{}{}{}{}\!\!\!\!\!\!\velopertXshort{}{0}{}{0}+\dots\\
         &=\barepropX{1}{1}{}{}\!\!\!\!\!\!\velopertXshort{}{0}{}{0} \sum_{k=0}^\infty\big(\barepropX{ 1}{ 1}{}{}\!\!\!\!\!\!\fricpertXshort{}{1}{}{1}\big)^k \\
         &=\effeV \frac{\Upsilon_{1,0}^{0,0}\effeV}{\gamma+\rotDiffusion}\frac{1}{1-\frac{\Lambda_{1,1}^{1,1}\effeV}{\gamma+\rotDiffusion}} +\mathcal{O}(\rescalefricTwoD^2)\ ,\elabel{ABP_phi11_final_2}
    \end{align}
\end{subequations}
By plugging \Eref{abp_vertices} and the dimensionless parameter \Eref{2d_dimensionless_para} into Eqs.~\eref{ABP_phi20_final_2} and \eref{ABP_phi11_final_2} , we obtain Eqs.~\eref{ABP_time_independent_observable_2} and \eref{ABP_time_independent_observable_3}, respectively.

\section{Numerical simulation}
\label{app:Numerical}
In this paper, we use the Euler-Maruyama method to simulate the system via the corresponding Langevin equation \Eref{Langevin_general},
\begin{subequations}
    \begin{align}
        \vvec_{i+1}-\vvec_i&=-\gamma \vvec_i \Delta t-\friction \frac{\vvec_i}{||\vvec_i||} \Delta t+\gamma \wvec_{\theta_i}\Delta t+ \sqrt{2D\gamma ^2 \Delta t} \xivec_i\ , \\
        \theta_{i+1}-\theta_i&=\sqrt{2\rotDiffusion \Delta t} \zeta_i\ ,\\
        \xvec_{i+1}-\xvec_i&=\vvec_i\Delta t \ ,
    \end{align}
\end{subequations}
with a suitable initial condition.
In the above, $\xivec_i$ is a 2D vector and the components are Gaussian random variables with zero expected value and unit variance. Similarly, $\zeta_i $ are also Gaussian random variables with zero expected value and unit variance. The effective diffusion coefficient is obtained from the mean-squared displacement
\begin{equation}
    \effeD^{(2)} = \frac{1}{4 N  M\Delta t} \sum_{i=1}^N (\xvec_M^{(i)}-\xvec_0^{(i)})^2 \ ,
\end{equation}
where $M$ is the total steps and $M\Delta t=T$ is the total time, and $N$ is the number of the particles  to obtain the ensemble average.

\end{document}

%% file: macros.tex
\newcommand{\epsilonvec}{\bm{\epsilon}}

\newcommand{\ave}[1]{\left\langle #1 \right\rangle}

\newcommand{\imag}{\mathring{\imath}}

\newcommand{\plaind}{\mathrm{d}}
\newcommand{\dint}[1]{\mathchoice{\!\plaind#1\,}{\!\plaind#1\,}{\!\plaind#1\,}{\!\plaind#1\,}}

\newcommand{\ddintx}[2]{\mathchoice{\!\plaind^{#2}#1\,}{\!\plaind^{#2}#1\,}{\!\plaind^{#2}#1\,}{\!\plaind^{#2}#1\,}}
\newcommand{\dTWOint}[1]{\ddintx{#1}{2}}

\newcommand{\dbar}{\plaind\mkern-6mu\mathchar'26}
\newcommand{\deltabar}{\delta\mkern-6mu\mathchar'26}

\newcommand{\Dint}[1]{\mathcal{D}\!#1\,}

\usepackage{dsfont}

\newcommand{\canetset}[1]{{\mathchoice {\hbox{$\sf\textstyle #1\kern-0.4em #1$}}
{\hbox{$\sf\textstyle #1\kern-0.4em #1$}}
{\hbox{$\sf\scriptstyle #1\kern-0.3em #1$}}
{\hbox{$\sf\scriptscriptstyle #1\kern-0.2em #1$}}}}

\def\nbZ{{\mathchoice {\hbox{$\sf\textstyle Z\kern-0.4em Z$}}
{\hbox{$\sf\textstyle Z\kern-0.4em Z$}}
{\hbox{$\sf\scriptstyle Z\kern-0.3em Z$}}
{\hbox{$\sf\scriptscriptstyle Z\kern-0.2em Z$}}}}

\newcommand{\gpvec}[1]{\mathbf{#1}}

\newcommand{\zerovec}{\gpvec{0}}
\newcommand{\nullvec}{\zerovec}

\newcommand{\vvec}{\gpvec{v}}
\newcommand{\wvec}{\gpvec{w}}
\newcommand{\xvec}{\gpvec{x}}

\usepackage{bm}
\newcommand{\xivec}{\bm{\xi}}

\newcommand{\Xivec}{\bm{\Xi}}

\newcommand{\ident}{\mathds{1}}

\newcommand{\AC}{\mathcal{A}}

\newcommand{\utilde}{\tilde{u}}

\newcommand{\phitilde}{\tilde{\phi}}

\renewcommand{\exp}[1]{\mathchoice{\mathrm{e}^{#1}}{\operatorname{exp}\left(#1\right)}{\operatorname{exp}\left(#1\right)}{\operatorname{exp}\left(#1\right)}}

\newcommand{\elabel}[1]{\label{eq:#1}}
\newcommand{\eref}[1]{(\ref{eq:#1})}
\newcommand{\Eref}[1]{Eq.~(\ref{eq:#1})}

\newlength \standardfigwidth
\newcounter{exercise}
{\addtocounter{exercise}{1}\begin{center}\begin{minipage}{0.8\linewidth}\textbf{Exercise
\arabic{exercise}:}\begin{itshape}}
{\end{itshape}\end{minipage}\end{center}}

\makeatletter
\newcommand{\creat}[3][]{\@ifempty{#1}{#2^{\dagger}}{\left(#2^{\dagger}\right)^{#1}}\@ifempty{#3}{}{\!(#3)}}

\newcommand{\creatDoi}[3][]{\@ifempty{#1}{\tilde{#2}}{\left(\tilde{#2}\right)^{#1}}\@ifempty{#3}{}{(#3)}}

\newcommand{\annih}[3][]{#2\@ifempty{#1}{}{^{#1}}\@ifempty{#3}{}{(#3)}}

\makeatother

% ...

\newlength{\bibmarkkeyAleft}

\newlength{\bibmarkkeyBleft}

\newlength{\bibmarkkeyCleft}

\newlength{\bibmarkkeyDleft}

\newcommand{\action}{\AC}
\newcommand{\actionPert}{\action_{\text{\tiny pert}}}